\documentclass[epj]{svjour}
\usepackage{graphicx}
\usepackage{amsmath}
\usepackage{ifthen}
\usepackage{amssymb}
\usepackage{color}
\newcommand{\slaninacolor}{false}
\ifthenelse{\equal{\slaninacolor}{true}}{%
}{
\renewcommand{\color}[1]{{}}

}
\begin{document}
\title{%
Analytical results for the Sznajd model of opinion formation
}%
\author{%
Franti\v{s}ek Slanina
\inst{1}\thanks{e-mail: {\tt slanina@fzu.cz}}
\and
Hynek Lavi\v{c}ka
\inst{2}
}%
\institute{
Institute of Physics,
 Academy of Sciences of the Czech Republic,\\
 Na~Slovance~2, CZ-18221~Praha,
Czech Republic%
\and
Faculty of Nuclear Sciences and Physical Engineering, Czech Technical 
 University in Prague,\\
B\v rehov\'a~7, CZ-11519~Praha 1,
Czech Republic%
}%
\date{%
}
%
\abstract{%
The Sznajd model, which describes opinion formation and social
influence, is treated analytically on a complete graph. We
prove the   existence of the phase transition in the original
formulation of the model, while for the Ochrombel modification we find
smooth behaviour without transition. We calculate the average time to
reach the stationary state as well as the exponential tail of its
probability distribution.  An analytical argument for the observed
$1/n$ dependence in the distribution of votes in Brazilian elections
is provided.
\PACS{
{89.65.-s}{Social and economic systems}\and
{05.40.-a}{Fluctuation phenomena, random processes, noise, and Brownian motion}\and
{02.50.-r}{Probability theory, stochastic processes, and statistics   }%
} 
} 
\maketitle
%
%
%

%
\section{Introduction} 
There is significant convergence between statistical physics and
mathematical sociology in approaches to their respective fields
\cite{schweitzer_02}. Ising model, the single most studied statistical
physics model, finds its numerous applications in sociophysics
simulations. Conversely, sociologically inspired models pose new
challenges to statistical physics. We believe this is the case of the
Sznajd model we are studying here.

The model of K. Sznajd-Weron and J. Sznajd \cite{szn_szn_00} was
designed to explain certain features of opinion dynamics. The slogan
``United we stand, divided we fall'' lead to simple dynamics, in which
individuals placed on a lattice (one-dimensional in the first version)
can choose between two opinions (political parties, products etc.) and
in each update step a pair of neighbours sharing common opinion
persuade their neighbours to  join their opinion. Therefore, it was
noted that contrary to the Ising or voter \cite{liggett_99} models,
information does not  flow from the neighbourhood to the selected
spin, but conversely, it flows out from the selected cluster to its
neighbours.

The model initiated a surge of immediate interest
\cite{ber_sta_ker_02,ber_cos_ara_sta_01,stauffer_02,stauffer_02a,stauffer_02b,stauffer_02c,mor_and_sta_01,sta_sou_mos_00,sta_deo_02,sta_deo_02a,stauffer_01,elgazzar_01,elgazzar_02,sznajd_02,szn_wer_02a,schulze_02,szn_wer_02,schulze_02a,ochrombel_01,bonnekoh_03,sab_ric_03,stauffer_03}
and the results of numerical simulations can be briefly summarised as
follows.  The results do not depend much on the spatial dimensionality
or on the type of the neighbourhood selected \cite{sta_sou_mos_00}.
In the case of $q$ choices of opinion, the system has $q$ obvious
homogeneous stationary (absorbing) states, where all individuals
choose the same opinion. There is no way to go out of the homogeneous
state, so it is an attractor of the dynamics.  This is reminiscent of
a zero-temperature dynamics, which in Ising model leads to rich
behaviour \cite{spi_kra_red_01}. However, in the Sznajd model, the
possible metastable states, like the ``antiferromagnetic''
configuration have negligible probability to occur, unless we
introduce explicitly also an ``antiferromagnetic'' dynamic rule as it
was used in   the very first formulation
\cite{szn_szn_00}.

The case $q=2$ was studied mostly, denoting the opinions by Ising
variables $+1$ and $-1$. The probability of hitting the stationary
state of all $+1$ (or, complementary,  all $-1$) was studied,
depending on the initial fraction $p$ of the  individuals choosing
$+1$. Sharp transition was observed at value $p=0.5$
\cite{sta_sou_mos_00}; for $p>0.5$ the probability to reach eventually
the state of all opinions $+1$ is close to one, while for $p<0.5$ it
is negligible, which can be interpreted as a dynamical phase
transition. The distribution of times  needed to reach the stationary
state was measured, revealing a peak followed by relatively fast
decay. This means that the average hitting time is a well-defined
quantity \cite{sta_sou_mos_00}.

It was also found in one and two-dimensional lattices that the
fraction of individuals who never changed opinion decays as a power
with time, similarly to Ising model. While the exponent in one
dimension agrees with the Ising case, the two-dimensional Sznajd model 
 gives different exponent than Ising model, indicating
different dynamical universality class \cite{sta_deo_02a}. Also the
waiting time between two subsequent opinion changes is  distributed
according to a power-law \cite{szn_szn_00}.

Among other studies, let us mention the influence of advertising
effects \cite{szn_wer_02a,schulze_02} and price formation
\cite{szn_wer_02}.  Long-range interactions were studied in
\cite{schulze_02a}.

In a very short but intriguing note  \cite{ochrombel_01} Ochrombel
suggested a drastic simplification of the Sznajd model. In the
Ochrombel version it is not necessary to have a cluster of identical
opinions. Any individual is capable to convince her neighbours to
select the same opinion. This model was reported to share all
essential features of the original Sznajd model, only the phase
transition in the probability of hitting the state of all $+1$ at
$p=0.5$ is absent.

The Sznajd model was also used to model the election process. There is
recent empirical evidence from  Brazilian elections
\cite{cos_alm_and_mor_99,lyr_cos_cos_and_02,cos_alm_mor_and_02}
that the distribution of votes per candidate follows a power-law, more
specifically $P(n)\sim 1/n$, where $n$ is the number of votes. This
result was reproduced in a study \cite{ber_sta_ker_02} based on Sznajd
model on a scale-free  network \cite{bar_alb_99,alb_bar_01,dor_men_01}.

The dynamics of elections was thoroughly investigated by S. Galam
\cite{galam_99,galam_00,galam_02,galam_03}, showing that majority rule
applied on sufficiently many hierarchical levels leads to a
homogeneous ``totalitarian'' state with one opinion pervading the
whole system. 

Other approaches to physical modelling of opinion dynamics were also
investigated
\cite{benn_kra_red_02,vaz_kra_red_02}
and among them especially the Axelrod model, which was found to have
rich behaviour from the statistical physics point of view
\cite{cas_mar_ves_00a,vil_ves_cas_02,kle_egu_tor_mig_03}.

We should also mention the well studied voter model
\cite{liggett_99,benna_fra_kra_96,dor_cha_cha_hin_01,mobilia_03},
which is very similar in spirit to the Sznajd model.  Indeed, the
relation of the two models was studied e. g. in
\cite{beh_sch_03} and it seems that Sznajd model reduces to the voter 
model at least for certain setups (especially using the Ochrombel
simplification on a complete graph) while for others the voter model
can be generalised so that it includes the rules of Sznajd model as a
special case.   In fact, similar analysis to that presented here was
performed for voter model, contact process and related processes in
\cite{dic_vid_02}. The persistence properties of the voter model on
complete graph were studied in \cite{benna_fra_kra_96}.

Very recently a ``Majority rule'' model, sharing some features with
Sznajd model, was introduced and studied in
\cite{kra_red_03} and its generalisation to the Majority-Minority
model \cite{mob_red_03} gives in the mean-field approximation results
closely related to ours.

\section{Formulation of the model and its simplifications}

\subsection{General scheme}

In the original formulation of the Sznajd model, the ``united we
stand'' principle is often stressed \cite{szn_szn_00,sta_sou_mos_00}.
It means that only a cluster of identical opinions can spread the same
opinion toward its neighbours. However, this principle was relaxed in
the Ochrombel simplification \cite{ochrombel_01} without qualitatively
affecting many of the results (except the presence of the phase
transition). We will propose some other simplifications here,
supposing the results remain robust.

Let us have $N$ agents, each of which can be in one of $q$ states
(opinions) $\sigma\in S$. We may for example think of a $q$-state
Potts model variables. Each agent sits on a node of a social network,
and they can interact along the edges with their nearest neighbours.

The opinion of the agent $i$ is denoted $\sigma_i$.  The state of the
system is described by the set of opinions of all the agents,
$\Sigma=[\sigma_1,\sigma_2,...,\sigma_N]$.

The variable $\Sigma(t)$ performs a discrete-time Markov process,
whose transition probabilities from time $t$ to $t+1$ differ in
various cases, which will be specified in the following.

\subsection{Case I: two against one}
\label{sec:caseI}
The first case investigated, which we will sometimes call ``two
against one'', generalises and simultaneously simplifies the various
versions introduced in \cite{sta_sou_mos_00}. The main difference is
in the fact that we will change at maximum {\it one} agent at each
time step. This may not significantly change the behaviour, as the
various  choices of neighbourhood in \cite{sta_sou_mos_00} exhibit
only little difference.

Our algorithm will iterate the following three steps. First, choose
randomly an agent $i$. Then, choose randomly one of its neighbours,
say $j$. If $\sigma_i(t)\ne\sigma_j(t)$, nothing happens. However, if
$\sigma_i(t) = \sigma_j(t)$, we will choose randomly one of the common
neighbours of both $i$ and $j$, say $k$, and set
$\sigma_k(t+1)=\sigma_i(t)$.  We may also write it schematically as
reactions $\mathrm{AAB}\to\mathrm{AAA}$,
$\mathrm{BBA}\to\mathrm{BBB}$.

\subsection{Case II: Ochrombel simplification}
\label{sec:caseII}
In this case, we do not need to have two neighbours in the same
state. Everybody can influence each of its neighbours.
We choose an agent $i$ at
random. Then, choose $j$ randomly among neighbours and
set $\sigma_j(t+1)=\sigma_i(t)$. 
Therefore, the process may be written as 
$\mathrm{AB}\to\mathrm{AA}$, $\mathrm{BA}\to\mathrm{BB}$.
In fact, on fully connected network the Ochro\-mbel simplification is
equivalent to voter model, whose dynamical properties were studied
e. g. in \cite{benna_fra_kra_96}.

As an obvious observation we can note that both in case I and case II
the uniform states, with all $\sigma_i$ equal, are stable under the
dynamics. However, we can expect variety of metastable states in the
case I, in which there are no pairs of neighbours in the same state,
therefore the dynamics does not proceed any further.

\section{On a fully-connected network}
\label{sec:meanfield}

We will approximate the complex social network by the fully-connected
network (the complete graph) of $N$ nodes. Here, any two  
agents are neighbours; in the case I we simply choose three agents
$i,j,k$ at random and in the case II two agents $i,j$ at random. Note
that the order in which they are chosen matters. This makes our
process different e. g. from the majority \cite{kra_red_03} or
majority-minority \cite{mob_red_03} models, although on fully
connected network the difference may consist only in rescaling certain
variables. 

We will call this setup a mean-field approximation in the same sense
as the Ising model on the complete graph can be considered as an
approximation for Ising model on hypercubic lattice of high
dimensionality. Of course this is not a good approximation to the
original one-dimensional formulation of the Sznajd model
\cite{szn_szn_00},
but we believe it is appropriate for much more realistic studies of
Sznajd model on complex networks
\cite{ber_sta_ker_02,elgazzar_02,bonnekoh_03}.  We refer the reader to
\ref{app:onarbitrarynetwork} for a more formal definition of
the Sznajd model on an arbitrary graph.

In fully-connected network the state of the system is fully described
by the occupation numbers $N_\sigma=\sum_{i=1}^N
\delta_{\sigma_i\sigma}$, or equivalently the densities
$n_\sigma=N_\sigma/N$, for each opinion $\sigma\in S$. The dynamics of
these occupation numbers fully describes the evolution of the
system. As the total number of nodes is conserved, there are $q-1$
independent dynamical variables.

Let us start with the case II (Ochrombel simplification) with only two
opinions, $q=2$. The variable $\sigma$ can assume only two values,
denoted $\sigma=\pm 1$ for convenience. Indeed, we are effectively working
with Ising spins. The state is described by one dynamical variable
only, which will be taken as a ``magnetisation'', 
\begin{equation}
m=\frac{N_+-N_-}{N}\quad .
\end{equation}
In one step of the dynamics, three events can happen. The
magnetisation may remain constant or it can change by $\pm 2/N$. The
probabilities of these three events can be easily calculated
\begin{eqnarray}
&{\rm Prob}&\left\{ m\to m +
\frac{2}{N}\right\}=
\frac{1}{4}\left(1-m^2\right)\left(1+\frac{1}{N-1}\right) \nonumber\\
&{\rm Prob}&\left\{ m\to m - \frac{2}{N}\right\}=
\frac{1}{4}\left(1-m^2\right)\left(1+\frac{1}{N-1}\right)
\label{eq:transitionprobabilitiesII}\\
&{\rm Prob}&\left\{ m\to m\right\}=
\left(\frac{1}{2}\left(1+m^2\right)-\frac{1}{N}\right)
\left(1+\frac{1}{N-1}\right)\nonumber\; .
\end{eqnarray}

Our objective is writing the master equation for the probability
density of the random variable $m(t)$, which we denote $P_m$. It can be
found easily in the 
thermodynamic limit $N\to\infty$. Indeed, we find that the time should
be rescaled as
\begin{equation}
t=N^2\, \tau
\label{eq:scalingII}
\end{equation}
in the thermodynamic limit. Then the probability density evolves
according to the partial differential equation
\begin{equation}
\frac{\partial}{\partial\tau}P_m(m,\tau)
=
\frac{\partial^2}{\partial m^2}\left[
(1-m^2)\,P_m(m,\tau)\right]\quad .
\label{eq:caseIIqis2}
\end{equation}
The latter equation describes in principle fully the evolution of the
Sznajd model in Ochrombel simplification on a complete graph.
 It has the form of a diffusion equation with
position-dependent diffusion constant.

Let us turn now to the case I (original Sznajd model), again with
$q=2$. We may repeat step by step the considerations made above for
the case II. Namely, our dynamical variable will be again the
magnetisation $m$ which may either remain unchanged or change by $\pm
2/N$ in one step. For the probabilities of these events we can find formulae
analogous to (\ref{eq:transitionprobabilitiesII})
\begin{eqnarray}
&&{\rm Prob}\left\{ m\to m +
\frac{2}{N}\right\}=
\frac{\left(1-m^2\right)}{8}\left(1+m 
+
\frac{1+3\,m}{N}
\right)
 \nonumber
\\
&&\begin{split}
{\rm Prob}\left\{ m\to m - \frac{2}{N}\right\}=
\frac{\left(1-m^2\right)}{8}\bigg(1&\,-m\, 
+
\\
&+
\frac{1-3\,m}{N}
\bigg)
\end{split}
\\
&&{\rm Prob}\left\{ m\to m\right\}=
1-\frac{\left(1-m^2\right)}{4}\left(1 
+
\frac{1}{N}
\right)
\nonumber
\label{eq:transitionprobabilitiesI}
\end{eqnarray}
where the terms of order $1/N^2$ are neglected. 
Note that the probabilities of changes $\pm 2/N$ are not
symmetric, contrary to the previous case (II). This fact has
all-important consequences. We will see later that it is responsible
for the fact that the original Sznajd model exhibits phase transition,
while in Ochrombel simplification the transition is absent. 

A more immediate consequence is that the time must be rescaled
differently, in order to get sensible thermodynamic limit, namely
\begin{equation}
t=2N\, \tau\quad .
\label{eq:scalingI}
\end{equation}
The second consequence is that the equation for $P_m(m,\tau)$ contains
{\it first} derivative with respect to $m$, representing a pure drift
in magnetisation:   
\begin{equation}
\frac{\partial}{\partial\tau}P_m(m,\tau)
=-
\frac{\partial}{\partial m}\left[
(1-m^2)m\,P_m(m,\tau)\right]\quad .
\label{eq:caseIqis2}
\end{equation}
Contrary to the previous case (\ref{eq:caseIIqis2}) the diffusion term,
containing the second derivative in $m$, represents only the
finite-size correction to the drift term. However, this correction may
dominate close to points $m=\pm 1$ and $m=0$ where the drift
velocity becomes zero.

Next case investigated will be the case II with arbitrary value of
$q$. Moreover, we will assume that the number of opinions is large, $q\gg 1$.
Let us define the distribution of occupation numbers
\begin{equation}
D(n)=\frac{N}{q}\sum_{\sigma=1}^q \delta\left(n-n_\sigma\right)
\end{equation}
where $\delta(x)=1$ for $x=0$ and zero elsewhere. It would be much more
difficult to write the full dynamic equation for $D(n)$. Therefore, we
use the approximation which replaces the distribution $D(n)$ by its
configuration average $P_n(n)=\langle D(n)\rangle$. In the limit
$N\to\infty$ and $q\to\infty$ and substituting the variable $x=2n-1$
we arrive at the equation
\begin{equation}
\frac{\partial}{\partial\tau}P_n(x,\tau)
=
\frac{\partial^2}{\partial x^2}\left[
(1-x^2)\,P_n(x,\tau)\right]\quad .
\label{eq:caseIIqislarge}
\end{equation}
The time is rescaled again according to the Eq. (\ref{eq:scalingII}).
We can see that the equations (\ref{eq:caseIIqis2}) and
(\ref{eq:caseIIqislarge}) have identical form, although the
interpretation of variables is different. We can therefore solve the
two cases simultaneously. This will be performed in the next section.

\section{Solution of the dynamics}

\subsection{Two against one: case I}
The case I, $q=2$ is described by the equation
\begin{equation}
\frac{\partial}{\partial\tau}P(x,\tau)
=-\frac{\partial}{\partial x}\left[(1-x^2)x\,P(x,\tau)\right]\; .
\label{eq:caseI}
\end{equation}
It can be easily verified that the solution has the following general form
\begin{equation}
P(x,\tau)
=[(1-x^2)x]^{-1}\,f({\rm e}^{-\tau}\,\frac{x}{\sqrt{1-x^2}})
\label{eq:solutioncaseI}
\end{equation}
for arbitrary  function $f(y)$. The form of the function $f(y)$
is given by initial conditions. For example if the initial condition
is a $\delta$-function, it keeps the same form during the evolution,
only the location shifts in time. This way we could in principle
calculate, how long it takes to reach the edges of the interval from
given initial position. This would be the time to reach the stationary
state. However, it comes out that the time needed
blows up. The reason comes from the infinite-size limit $N\to
\infty$. Indeed, very close to the points $x=\pm 1$ the finite-size
effects take over. 

We can estimate the average time needed to
reach the stationary state in finite system by the following
consideration. In fact, 
the equation (\ref{eq:caseI}) describes the drift which pushes the
system toward the stationary state, but neglects the effect of
diffusion, which becomes important at a distance $\sim 1/N$ from the
points $x=\pm 1$. Therefore, we must calculate the time necessary for
the drift to drive the system to the point $\pm(1-1/N)$. The initial
fraction $p$ of opinions $+1$ corresponds to the initial condition $x_0=2p-1$
and from the formula (\ref{eq:solutioncaseI}) we have the following
estimate for the average time $\langle\tau_{\rm st}\rangle$ to reach
the stationary state 
\begin{equation}
\langle\tau_{\rm st}\rangle
\simeq
-\ln\left(\frac{|2p-1|}{\sqrt{p(1-p)}}\,\frac{1}{\sqrt{N}}\right)\; .
\label{eq:avwaitingtimecaseI}
\end{equation}

It is also possible to include the correction terms of order $O(1/N)$
into Eq. (\ref{eq:caseI}) and deduce the equation for the average time
to reach the absorbing state $\langle\tau_{\rm st}\rangle(x_0)$ on
condition that the process started at initial position $x_0$. Following
the general scheme \cite{gardiner_85} we obtain a second-order
ordinary differential equation
\begin{equation}
\begin{split}
\left( 1+\frac{3}{N} \right) \left( 1-{x_0}^{2} \right)x_0\,&   {\frac {\mathrm{d}}{\mathrm{d}x_0
}}\langle\tau_{\rm st}\rangle \left( x_0 \right) +\\
+\frac {1}{N} \left( 1-{x_0}^{2} \right)\,& \frac {\mathrm{d}^{2}}
{\mathrm{d}{x_0}^{2}} \langle\tau_{\rm st}\rangle\left( x_0 \right)
=-1\;.
\end{split}
\label{eq:foravwaitingtimecaseI}
\end{equation}
The solution of (\ref{eq:foravwaitingtimecaseI})  is
\begin{equation}
\langle\tau_{\rm st}\rangle \left( x_0 \right)=
N\int_{-1}^{x_0}\int_y^0\frac{\mathrm{e}^{\frac{N+3}{2}\,z^2}}{1-z^2}\mathrm{d}z
\;\mathrm{e}^{-\frac{N+3}{2}\,y^2}\,\mathrm{d}y\; .
\label{eq:solutionforavwaitingtimecaseI}
\end{equation}
Indeed, for $x_0$ not too close to either of the points $x_0=-1,0,1$ (the
distance must be large compared to $1/N$) we obtain from the formula
(\ref{eq:solutionforavwaitingtimecaseI}) an approximate expression of
the form given in (\ref{eq:avwaitingtimecaseI}). Another way to obtain
the same $p$ dependence as in (\ref{eq:avwaitingtimecaseI}) is to omit
the $O(1/N)$ terms in the equation (\ref{eq:foravwaitingtimecaseI})
and solve the first-order differential equation. In this case,
however, we lose any information about the dependence on $N$. We
should also note that a result essentially equivalent to
Eq. (\ref{eq:avwaitingtimecaseI})  was obtained also in \cite{kra_red_03}. 

It is rather interesting to observe that the deterministic dynamics of
Galam model \cite{galam_00,galam_03} leads to a formula very similar to
(\ref{eq:avwaitingtimecaseI}), while the interpretation of the time
variable is totally different: in Galam model it represents the number
of hierarchical levels on which the majority rule is iterated.

It would be desirable to calculate the full probability distribution
for the time to reach the stationary state  $\tau_{\rm st}$  and not
only the average. That is possible using again 
the formalism of adjoint equation \cite{gardiner_85}, when we
introduce the $1/N$ corrections 
to  Eq. (\ref{eq:caseI}) but the resulting partial differential
equation is difficult to solve explicitly.  Instead, we  estimate the
exponential tail of the distribution by a simple consideration.

Indeed, after the drift had pushed the system to the state in which
there is only single spin $-1$ immersed in a sea of all $+1$-s it
finally comes into uniform stationary state if the first pair of spins
chosen is both $+1$ and the third one is the single $-1$. This choice
has probability $\simeq 1/N$. Therefore, the relaxation time toward the uniform
state is $t_{\rm relax}\simeq N$ and using the scaling
(\ref{eq:scalingI}) we have for the tail of the distribution
\begin{equation}
P(\tau_{\rm st})\sim\exp(-\frac{\tau_{\rm st}}{\tau_{\rm
  relax}}),\quad\tau_{\rm 
  st}\to\infty
\label{eq:tailwaitingtimecaseI}
\end{equation}
with 
\begin{equation}
\tau_{\rm relax}\simeq\frac{1}{2}\quad .
\label{eq:relaxwaitingtimecaseI}
\end{equation}

The most important observation we can draw from the solution
(\ref{eq:solutioncaseI}) 
is the presence of the dynamic phase transition, as observed in
numerical simulations. Indeed, starting with any fixed positive
magnetisation, we have initial condition $P(x,0)=\delta(x-x_0)$,
$x_0>0$, and 
the drift expressed by Eq. (\ref{eq:solutioncaseI}) always take us to
the state with all agents having opinion  $+1$, while from  
any state with negative magnetisation the drift leads the system eventually 
to the state with all agents having opinion $-1$ 
and the probability of ending in the state of all $+1$ is therefore
$P_+=\theta(p-1/2)$.
The possible deviations from this
rule close to the zero magnetisation (i. e. $p=0.5$) are due to the
finite size effects, which are neglected in  
(\ref{eq:caseI}).
The presence of the phase transition is also indicated by the divergence
of the average time to reach the stationary state
(\ref{eq:avwaitingtimecaseI}) for $p\to 1/2$.

\subsection{Ochrombel simplification: case II}
The equation
\begin{equation}
\frac{\partial}{\partial\tau}P(x,\tau)
=\frac{\partial^2}{\partial x^2}\left[(1-x^2)\,P(x,\tau)\right]
\label{eq:caseII}
\end{equation}
describes both the case II, $q=2$ and II, $q\gg 1$, only the
interpretation of the variable $x$ differ: in the former case it corresponds
to the magnetisation, while in the latter case it is shifted
percentage of votes. By solving  Eq. (\ref{eq:caseII}) we treat
simultaneously both cases. 

The equation of the form (\ref{eq:caseII}) was already studied in
variety of contexts, e. g. population genetics 
\cite{wright_45,dor_lan_83} or
reaction kinetics \cite{bena_con_mea_red_tak_90} and can be tackled by
standard methods developed for Fokker-Planck equation.

Indeed, we look for the solution using the expansion in eigenvectors.
We can write (\ref{eq:caseII}) it in the form
$
\frac{\partial}{\partial\tau}P(x,\tau)
=\mathcal{L}P(x,\tau)
$
where the linear operator $\mathcal{L}$ acts as
$
(\mathcal{L}f)(x)=
\frac{\partial^2}{\partial x^2}\left[(1-x^2)\,f(x)\right]
$.
We therefore need to find the set of eigenvectors of
$\mathcal{L}$. Denoting $\Phi_c(x)$ the eigenvector corresponding to
the eigenvalue $-c$, we have the following equation  
\begin{equation}
(1-x^2)\,\Phi_c''(x)-4x\,\Phi_c'(x)+(c-2)\,\Phi_c(x)=0\; .
\label{eq:eigenvectors}
\end{equation}
The full solution of (\ref{eq:caseII}) can be then expanded as
\begin{equation}
P(x,\tau)=\sum_c A_c{\rm e}^{-c\tau}\,\Phi_c(x)
\label{eq:expansionineigenvectors}
\end{equation}
with coefficients $A_c$ determined from the initial condition.

Important question to be settled prior to the attempt for solution is,
what is the appropriate space of functions $\Phi(x)$. First, the
interpretation of these functions  as probability densities sets the
requirement that it must be normalisable: $\int \Phi(x)\,{\rm d}x<\infty$.
Second, only the interval $x\in [-1,1]$ is relevant, so $\Phi(x)=0$ outside
this interval. Finally, we should anticipate the possibility that
$\delta$-functions appear in the solution, namely located at
$x=\pm 1$, because the uniform states, with all sites carrying the same
spin value, are stable under the dynamics.

We therefore look for the solution of  (\ref{eq:eigenvectors})
in the space of distributions (i.e. linear functionals on sufficiently
differentiable functions) with support restricted to the interval $[-1,1]$.

It is straightforward to find the eigenvectors corresponding to
eigenvalue $c=0$, i. e. the stationary solutions of the equation
(\ref{eq:caseII}). They are composed of $\delta$-functions only.
In fact, the corresponding eigensubspace is
two-dimensional and the base vectors can be chosen as
\begin{equation}
\Phi_{01}=\delta(x-1)\quad,\quad \Phi_{02}=\delta(x+1)\quad .
\label{eq:phizero}
\end{equation}

For $c\ne 0$ we first decompose the solution in ordinary function
of $x$ plus  a pair of $\delta$-functions, namely
\begin{equation}
\begin{split}
\Phi_{c}=\phi_{c+}\,&\delta(x-1)+\phi_{c-}\,\delta(x+1)+\\
&+\phi_c(x)\,\theta(x-1)\,\theta(x+1)  
\label{eq:Phidecomposition}
\end{split}
\end{equation}
where 
$\phi_{c+}$ and $\phi_{c-}$ are real numbers and $\phi_c(x)$ is a real
doubly differentiable function. 
Then, the equation (\ref{eq:eigenvectors}) translates into equation
for $\phi_c(x)$ 
\begin{equation}
(1-x^2)\,\phi_c''(x)-4x\,\phi_c'(x)+(c-2)\,\phi_c(x)=0
\label{eq:forphi}
\end{equation}
accompanied by two other conditions
\begin{equation}
\lim_{x\to\pm 1}\phi_{c}(x)=-\frac{c}{2}\,\phi_{c\pm}\quad .
\label{eq:phiplusminus}
\end{equation}

The general solution of the equation (\ref{eq:forphi}) exhibits
behaviour $\phi_c(x)\sim (1\mp x)^\alpha$ at $x\to\pm 1$, where either
$\alpha=0$ or $\alpha=-1$. However, the latter case should be
excluded, as it gives non-normalisable probability distribution. In
fact it is the condition of normalisability that  determines   all
possible eigenvalues $c$. The solution of (\ref{eq:forphi}) with
correct behaviour at $x\to\pm 1$ can be
expressed in Gegenbauer polynomials
\cite{bena_con_mea_red_tak_90,redner_01,gra_ryz_94}. The eigenvalues
are $c=c_l\equiv (l+1)(l+2)$ for $l=0,1,2,...\,$. An elementary
solution and the table of several lowest polynomials is presented in
\ref{app:expansionineigenvectors}.  

It is important to note that for any eigenvalue $c>0$ we have
\begin{equation}
\int\Phi_c(x)\,{\rm d}x = 0\qquad \int x\,\Phi_c(x)\,{\rm d}x = 0\quad .
\label{eq:momentsarezero}
\end{equation}
The consequence is that both $\int P(x,\tau){\rm d}x$ and $\int
x P(x,\tau){\rm d}x$ are independent of time. While the first
conservation law
expresses simply the conservation of probability, the second one is a
non-trivial consequence of the model dynamics. 
Mathematically it is related to the fact that the eigenspace
corresponding to zero eigenvalue is two-dime\-nsional.

Thus, we found the set of {\it right} eigenvectors of the operator
$\mathcal{L}$. 
For practical solution we still need to establish the coefficients
$A_c$ in Eq. (\ref{eq:expansionineigenvectors}). To this end we need
also the set of {\it left} eigenvectors of $\mathcal{L}$, checking
simultaneously that the set of left and right eigenvalues
coincide. First, we need to establish the adjoint operator to
$\mathcal{L}$, defined by usual relation
$(\mathcal{L}f|g)=(f|\mathcal{L}^T g)$. While $\mathcal{L}$ acts on
the space of distributions, its adjoint $\mathcal{L}^T$ acts on the
corresponding dual space, which is the space of sufficiently
differentiable functions. Straightforward algebra gives
$
(\mathcal{L}^T g)(x)=(1-x^2)\,g''(x)
$
which implies the following equation for the left eigenvectors
\begin{equation}
(1-x^2)\,\psi_c''(x)+c\,\psi_c(x)=0\quad .
\label{eq:forpsi}
\end{equation}

We find again that for $c=0$ the eigensubspace is two-dimensional. We
can choose the basis vectors so that they are mutually ortho-normal to
the pair of right eigenvectors (\ref{eq:phizero}), namely
\begin{equation}
\psi_{01}=\frac{1}{2}(1+x)\quad,\quad \psi_{02}=\frac{1}{2}(1-x)\; .
\label{eq:psizero}
\end{equation}
The solutions of (\ref{eq:forpsi}) for $c>0$ with proper boundary
conditions are again polynomials presented in more detail in 
\ref{app:expansionineigenvectors}.

The coefficients in the solution (\ref{eq:expansionineigenvectors})
with initial condition $P(x,0)=P_0(x)$ are then calculated as
\begin{equation}
A_c=\frac{\int P_0(x)\psi_c(x)\,{\rm d}x}{\int
\phi_c(x)\psi_c(x)\,{\rm d}x}\quad .
\label{eq:coefficientsAc}
\end{equation}

From the solution (\ref{eq:expansionineigenvectors}) we can deduce an
important feature for the distribution of waiting times needed to reach the
stationary state. Indeed, if $P_{\rm st}(\tau)$ is the probability
density for ending at time $\tau$ in the stationary frozen
configuration with all agents in 
the same state, we can express the probability that the stationary
configuration was not reached before time $\tau$ as
\begin{equation}
\begin{split}
P^>_{\rm st}(\tau)&\equiv\int_{\tau}^\infty P_{\rm
st}(\tau'){\rm d}\tau'=\\
&=1-\lim_{\epsilon\to 0^+}
\left(\int_{-1-\epsilon}^{-1+\epsilon}+\int_{1-\epsilon}^{1+\epsilon}\right)
P(x,\tau){\rm d}x\; .
\end{split}
\end{equation}
We can see that only the $\delta$-function components of the
eigenvectors $\Phi_c(x)$ in the expansion 
(\ref{eq:expansionineigenvectors}) contribute to $P^>_{\rm
st}(\tau_{\rm st})$. More explicitly, we find
\begin{equation}
P^>_{\rm st}(\tau)=
\sum_{c>0} 2A_c\, \frac{\phi_{c}(-1)+\phi_{c}(1)}{c}\, {\rm
e}^{-c\tau}\; .
\label{eq:expansionofwaitingtime}
\end{equation}
As the spectrum of eigenvalues is discrete, for long
times only the lowest non-zero $c$ (equal to $c_0=2$) is relevant.
Therefore, the distribution of waiting times will have an exponential
tail 
$P^>_{\rm st}(\tau)\sim{\rm e}^{-2\tau},
\;\tau\to\infty$.
For initial condition $P_0(x)=\delta(x-x_0)$ we can easily compute also the
prefactor in the leading term for large $\tau$. Indeed, from
(\ref{eq:coefficientsAc}) we  get $A_2$ and finally obtain
\begin{equation}
P^>_{\rm st}(\tau)\simeq \frac{6}{4}(1-x_0^2)\,{\rm e}^{-2\tau},
\;\tau\to\infty\; .
\label{eq:waitingtimedistributiontail}
\end{equation}
As the functions $\phi_c(x)$ are odd for $c=c_l$ with odd $l$, 
we should expect that the corrections to the formula
(\ref{eq:waitingtimedistributiontail}) 
will be governed by the second next
eigenvalue $c_2=12$. We will see later how it can be checked in
numerical simulations.

As in the case I the average time 
$\langle\tau_{\rm st}\rangle \left( x_0 \right)$ 
to reach the absorbing state when
starting at position $x_0$ can be obtained, using the general formalism
 \cite{gardiner_85}, from the equation
\begin{equation}
\left( 1-{x_0}^{2} \right)\, \frac {\mathrm{d}^{2}}
{\mathrm{d}{x_0}^{2}} \langle\tau_{\rm st}\rangle\left( x_0 \right)
=-1
\label{eq:foravwaitingtimecaseII}
\end{equation}
which can be solved easily
\begin{equation}
\langle\tau_{\rm st}\rangle\left( x_0 \right)
=
-\frac{x_0}{2}\ln\frac{1+x_0}{1-x_0}-\frac{1}{2}\ln\frac{1-{x_0}^2}{4}
\label{eq:solutionforavwaitingtimecaseII}
\end{equation}
(see also \cite{bena_con_mea_red_tak_90,redner_01}). The method of
adjoint equation \cite{gardiner_85,redner_01} can be used to calculate
the distribution of times to reach the absorbing state, when starting
from initial position at $x=x_0$, yielding results equivalent to our
direct calculation. Indeed, inserting the initial
condition  $P_0(x)=\delta(x-x_0)$ into 
(\ref{eq:coefficientsAc}) we can see that the expression
(\ref{eq:expansionofwaitingtime}) represents an expansion in
the eigenvectors $\psi_c(x_0)$ of the adjoint operator
$\mathcal{L}^T$ taken at point $x_0$.

Contrary to the case I, we do not observe any phase transition
here. This is due to the conservation of average magnetisation in  the
dynamics \cite{kra_red_03}. 
From this fact it follows  immediately that $P_+=p$. This
result can be confirmed by an explicit calculation.  Starting with fixed
magnetisation $x_0=2p-1$, the initial condition $P(x,0)=\delta(x-x_0)$
broadens under the diffusive dynamics (\ref{eq:caseII}) and leaves
always non-zero probability of ending in either of the possible
stationary states.  We already noted that  $\int x\,P(x,\tau)\,{\rm
d}x$ is independent of time under the dynamics
(\ref{eq:caseII}). Therefore, the asymptotic state  is the following
combination of the eigenvectors (\ref{eq:phizero})  with $c=0$
\begin{equation}
\lim_{\tau\to\infty} P(x,\tau)=
\frac{1-x_0}{2}\,\delta(x+1)+
\frac{1+x_0}{2}\,\delta(x-1)
\end{equation}
and the probability of ending in the state of all $+1$ is therefore simply
$P_+=p$.

\begin{figure}[t]
\includegraphics[scale=0.8]{%
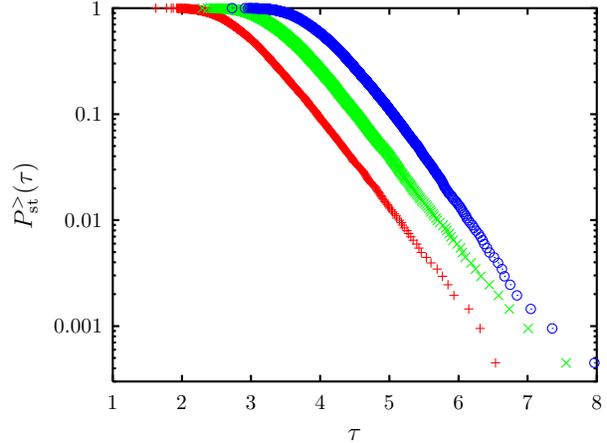}
\caption{%
Probability of reaching the stationary state in time larger than
$\tau$, for case I, $q=2$, $N=2000$. The values of initial fraction $p$ of
opinions $+1$ are 0.1
($+$) 0.2 ($\times$)
and 0.7 ($\odot$).
}
\label{fig:plarge-2vs1-2e3}
\end{figure}
\begin{figure}[t]
\includegraphics[scale=0.8]{%
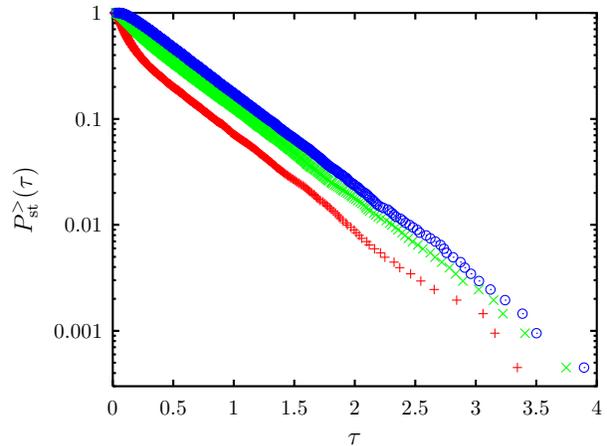}
\caption{%
Probability of reaching the stationary state in time larger than
$\tau$, for case II, $q=2$, $N=2000$. The values of initial fraction $p$ of
opinions $+1$ are 0.1
($+$) 0.2 ($\times$)
and 0.7 ($\odot$).
}
\label{fig:plarge-1vs1-2e3}
\end{figure}

\subsection{Distribution of votes}

As already stressed in Sec. \ref{sec:meanfield},  equation
(\ref{eq:caseII}) describes also the evolution of the distribution of
votes in the case of $q\gg 1$ parties. We will present an argument how
our results may explain the empirical data, suggesting the $1/n$ law
for the distribution of votes.

As stressed in the discussion following  Eq.
(\ref{eq:phiplusminus}), the time-independent solutions of
Eq. (\ref{eq:caseII}) can behave either as $1+x$ or $(1+x)^{-1}$ in
the limit $x\to -1$. However, the latter case was excluded by the
requirement of normalisability of the probability density. On the
other hand, relaxing the normalisability condition, the functions 
\begin{eqnarray}
\tilde{\phi}_{01}(x)&=&\frac{1}{1+x}
\label{eq:longtransientphione}\\
\tilde{\phi}_{02}(x)&=&\frac{1}{1-x}
\label{eq:longtransientphitwo}
\end{eqnarray}
are solutions of (\ref{eq:forphi}) with eigenvalue $c=0$. (Of course,
any linear combination of them is also solution with $c=0$). 

How should be any of these additional solutions interpreted? The zero
eigenvalue suggest that the function is stationary in time. However,
it is not normalisable, therefore this solution cannot be reached from
any initial condition. But if the distribution $P_n(x,\tau)$ is close
to $\tilde{\phi}_{01}(x)$ (or $\tilde{\phi}_{02}(x)$) 
in some interval $I$ of $x$, it is probable that it
$P_n(x,\tau)$ will remain close to (\ref{eq:longtransientphione}) (or
(\ref{eq:longtransientphitwo}), respectively) for
certain period of time, while the interval $I$ will gradually shrink and
eventually disappear. Therefore, we may suggest
(\ref{eq:longtransientphione}) and  (\ref{eq:longtransientphitwo}) as
a metastable states, or long-lived 
transient states.

This may explain the observation from simulations performed in
\cite{ber_sta_ker_02}. In this work, the distribution of the type $1/n$
is obtained in a suitably chosen transient regime, in certain range of
$n$. As $x=2n-1$, the behaviour of (\ref{eq:longtransientphione}) at
$x\to -1$ corresponds precisely to $1/n$ behaviour for small $n$.

A slightly more rigorous variant of the above argument is also
possible. Imagine 
now that the political system represented by the set of opinions $S$ is
not closed, but new opinions may appear, replacing other ones which
vanish. 

Indeed, the current induced by the dynamics of case II can be read off
from Eq. (\ref{eq:caseII})
\begin{equation}
j=-\frac{\partial}{\partial x}[(1-x^2)P(x,\tau)]
\end{equation}
and by insertion of the solution (\ref{eq:longtransientphione})
we deduce that there is homogeneous flow $j=+1$ outward the value $x=-1$,
i. e. $n=0$. We may interpret this flow as a consequence of an
external source placed somewhere close to the point $x=-1$,
i. e. $n=0$. Such a source accounts for the influx of new opinions, or
new parties, into the system. It is very reasonable to assume that the
source is placed at very small values of $n$, as new subjects are
likely to gain little support initially.

\begin{figure}[t]
\includegraphics[scale=0.8]{%
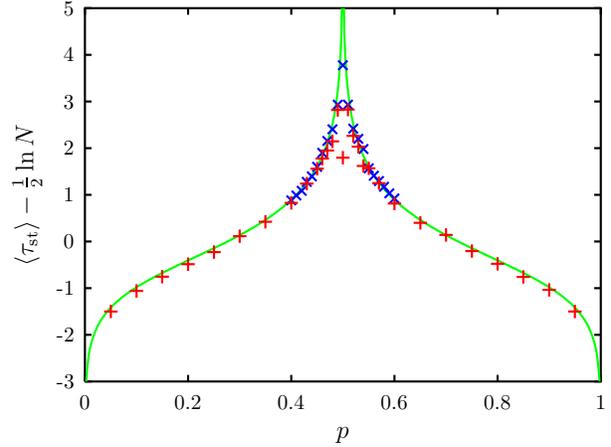}
\caption{%
Average time of reaching the stationary state in dynamics of case I, $q=2$. The
system size is $N=2000$ ($+$) and $N=4000$ ($\times$).  The line is
the analytic prediction of Eq. (\ref{eq:avwaitingtimecaseI}).  
}
\label{fig:avtau-2vs1-2e3}
\end{figure}
\begin{figure}[t]
\includegraphics[scale=0.8]{%
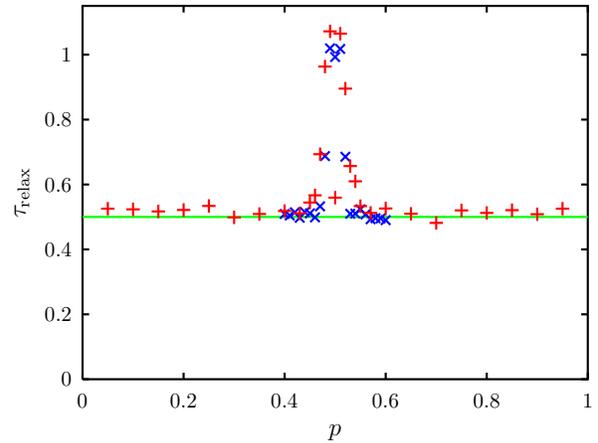}
\caption{%
Relaxation time toward the stationary state in dynamics of case I, $q=2$. The
system size is $N=2000$ ($+$) and $N=4000$ ($\times$).  The
horizontal line is the analytic prediction of
Eq. (\ref{eq:relaxwaitingtimecaseI}).  
}
\label{fig:trelax-2vs1-2e3}
\end{figure}

\section{Comparison with numerical simulations}

We performed numerical simulations of the Sznajd model on fully
connected network according to algorithms described in sections 
\ref{sec:caseI} (case I)
and \ref{sec:caseII} (case II). The main focus was on the dynamical
properties, namely the distribution of times needed to reach the
homogeneous stationary state.
We show in figures \ref{fig:plarge-2vs1-2e3} and
\ref{fig:plarge-1vs1-2e3} the probabilities $P^>_{\rm st}(\tau)$ that
the time $\tau_{\rm st}$ to reach the stationary state is larger that
$\tau$. We can clearly see that the probability decays exponentially
with $\tau$ in both cases I and II.

Let us discuss the case I first. Following the analytical expectation
(\ref{eq:tailwaitingtimecaseI}) we can fit the exponential tail of
the distribution as
\begin{equation}
P_{\rm st}^>(\tau)\simeq \exp\left(-\frac{\tau-\langle\tau_{\rm
st}\rangle
}{\tau_{\rm relax}}\right), 
\quad\tau\to\infty\quad .
\end{equation}
The results for $\langle\tau_{\rm
st}\rangle$ can be seen in Fig. \ref{fig:avtau-2vs1-2e3}, compared
with the analytical prediction of
Eq. (\ref{eq:avwaitingtimecaseI}). Similarly in
Fig. \ref{fig:trelax-2vs1-2e3} we can compare the fitted relaxation
time with the analytical result. Both $\langle\tau_{\rm st}\rangle$
and $\tau_{\rm relax}$ agree satisfactorily with the analytical
predictions. The deviations around the value $p=0.5$ are due to finite
size effects; the comparison of the results for system sizes $N=2000$
and $N=4000$ supports this interpretation.
From Eq. (\ref{eq:avwaitingtimecaseI}) we can see that
$\langle\tau_{\rm st}\rangle$ diverges logarithmically for
$N\to\infty$. This is confirmed by the simulation data which fall onto
single curve in Fig. \ref{fig:avtau-2vs1-2e3} for different system sizes.

Now let us turn to the case II.
The equation
(\ref{eq:expansionofwaitingtime})
yields the leading term in the
tail of the distribution $P^>_{\rm st}(\tau)$ and in principle 
also the corrections to it. As the functions $\phi_c(x)$ are odd for
$c=c_l$ with odd $l$, the next non-zero correction will come from the
eigenvalue $c_2=12$. Therefore, we expect the behaviour
\begin{equation}
P^>_{\rm st}(\tau)\simeq \exp\left(-\frac{\tau-\tau_0}{\tau_{{\rm
  r}0}}\right)+a_1\exp(-\frac{\tau}{\tau_{{\rm
  r}1}}),\;\tau\to\infty
\label{eq:tailwaitingtimecaseII}
\end{equation}
with 
\begin{equation}
\tau_{{\rm r}0}=\frac{1}{2},\quad \tau_{{\rm r}1}=\frac{1}{12}
\quad .
\label{eq:relaxtimescaseII}
\end{equation}

\begin{figure}[t]
\includegraphics[scale=0.8]{%
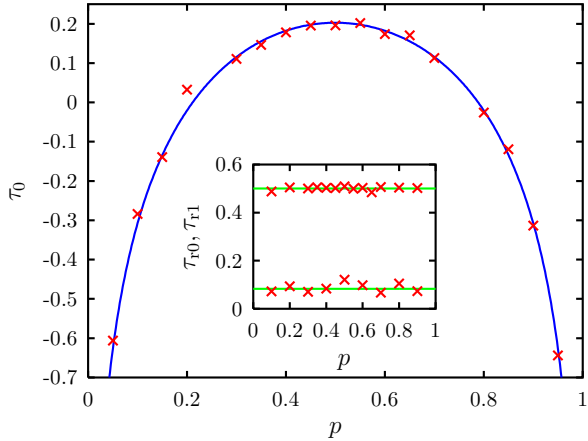}
\caption{%
The fitted parameter $\tau_0$ for reaching the stationary state in dynamics of
case II, $q=2$. The 
system size is $N=2000$. The line represents the formula
(\ref{eq:tau0caseII}). In the inset, the fitted first two relaxation times 
$\tau_{{\rm r}0}$ and $\tau_{{\rm r}1}$ are shown. 
The horizontal lines are corresponding
analytical predictions from Eq. (\ref{eq:relaxtimescaseII}).  
}
\label{fig:avtau-1vs1-2e3}
\end{figure}

As in the initial condition $P_0(x)=\delta(x-x_0)$ we have $x_0=2p-1$,
we can deduce from 
Eq. (\ref{eq:waitingtimedistributiontail}) the following estimate
\begin{equation}
\tau_0\simeq \ln\sqrt{6\,p\,(1-p)}\quad .
\label{eq:tau0caseII}
\end{equation}
We can see from Fig. \ref{fig:avtau-1vs1-2e3} that 
it corresponds well to the numerical data.
In the inset of Fig. \ref{fig:avtau-1vs1-2e3} we can also see the
fitted relaxation times $\tau_{{\rm r}0}$ and $\tau_{{\rm r}1}$. Also
here the correspondence with analytical prediction
(\ref{eq:relaxtimescaseII}) is good.

\section{Conclusions}

We formulated a mean-field version
of the Sznajd model of opinion formation by putting it on a complete
graph. Solving the underlying diffusion equations we found analytical 
results for several dynamical properties, as
well as exact long-time asymptotics.
 The results differ
substantially in the two cases studied: first, the original Sznajd
model, where a cluster of 
identical opinions is necessary to persuade others to join them, and
second,  the Ochrombel simplification, where also isolated agent can
persuade others. Dynamical phase transition was found analytically in the
original Sznajd model, while in the Ochrombel version it is
absent. This finding agrees with previous numerical results.

The approach to stationary state was the main concern of our
calculations. We found that the distribution of times to reach the
stationary state has an exponential tail which we were able to
calculate analytically. In the case of Ochrombel simplification, we
obtained also the corrections and a formula which gives in principle
the whole distribution. We compared the analytical results for the
tail (and in the Ochrombel case also for the first  correction) with
numerical simulations and we found good agreement. The method of
adjoint equation enabled us to find analytically 
the average time to
reach the stationary state, in both cases. 

We found also
another signature of the phase transition in the original Sznajd
model, expressed by the divergence of the average time to reach the
stationary state. Contrary to the Ochrombel case, in the original
Sznajd model the average time needed for reaching the stationary state
blows up logarithmically with increasing system size. This finding was
also confirmed in our numerical simulations.

The analytical treatment provided an explanation of the $1/n$
distribution of votes, documented in Brazilian elections. We found
that this behaviour corresponds to long-lived transient state of the
dynamics of the Sznajd model with large number of possible opinions, 
or alternatively to the dynamics of an
open version of the Sznajd model, where new opinions may continuously
emerge.

\begin{acknowledgement}
This work was supported by the project No. 202/01/1091
of the Grant Agency  of the Czech Republic. 
\end{acknowledgement}
\setcounter{section}{0}
\setcounter{equation}{0}
\renewcommand{\thesection}{Appendix \Alph{section}}
\renewcommand{\theequation}{\Alph{section}.\arabic{equation}}
\section{: Sznajd model on an arbitrary social network}
\label{app:onarbitrarynetwork}

Our system is composed of $N$ agents placed on nodes of a social
network, represented by the graph $\Lambda=(\Gamma,E)$ where $\Gamma$
is the set of nodes and $E$ set of edges, i. e. unordered pairs of
nodes. For a node $i\in\Gamma$ we denote
$\Gamma_i=\{j\in\Gamma|(i,j)\in E\}$ the set of neighbours of
$i$.

The opinion of the agent $i$ s denoted $\sigma_i$.
The state of the system is described by the set of opinions
of all the agents, $\Sigma=[\sigma_1,\sigma_2,...,\sigma_N]\in S^\Gamma$.
The variable $\Sigma(t)$ performs a discrete-time Markov process,
defined as follows.

In the case I we iterate the following three steps. First, choose
$i\in\Gamma$ at  random. Then, choose $j\in\Gamma_i$ randomly among
neighbours of $i$. If $\sigma_i(t)\ne\sigma_j(t)$, nothing
happens. However, if $\sigma_i(t) = \sigma_j(t)$, we will choose
randomly one of the common neighbours
$k\in\Gamma_i\cap\Gamma_j\setminus\{i,j\}$ and set
$\sigma_k(t+1)=\sigma_i(t)$.

In the case II we choose $i\in\Gamma$ at random. Then, choose
$j\in\Gamma_i$ randomly among neighbours and set
$\sigma_j(t+1)=\sigma_i(t)$.

If the graph is random and densely connected, we may approximate it
by the complete graph with $N$ nodes, i. e. for each pair of
nodes $i,j\in\Gamma$ there is an edge connecting them,  $(i,j)\in
E$. It means that the set of neighbours of a node $i\in\Gamma$ is
$\Gamma_i=\Gamma\setminus \{i\}$. This is a kind of a mean-field
approximation.

\section{: Finding the eigenvectors}
\label{app:expansionineigenvectors}

We can look for the solution of the equation (\ref{eq:forphi}) in the
form of power series
\begin{equation}
\phi_c(x)=\sum_{l=0}^\infty b_l\,x^l
\label{eq:seriesforphi}
\end{equation}
and find the recurrence relation for the coefficients
\begin{equation}
b_{l+2}      = \left(1-\frac{c}{(l+1)(l+2)}\right)\,b_l\; .
\end{equation}

We should distinguish two cases. Either the sequence of coefficients $b_l$
contains non-zero values for arbitrarily large $l$, or it is truncated
at some order and (\ref{eq:seriesforphi}) becomes a polynomial. In the
former case the solution behaves as $\phi_c(x)\sim (1-x^2)^{-1}$ at
$x\to\pm 1$ and  must be excluded. The latter case is possible only if
\begin{equation}
c=c_{l}      \equiv (l+1)(l+2)
\label{eq:eigenvalues}
\end{equation}
for some $l\ge 0$.
Moreover, in order to have a solution in the form of a polynomial, we
require that $b_1=0$ if $l$ in the equation (\ref{eq:eigenvalues}) is even, and $b_0=0$ if
$l$ in the equation (\ref{eq:eigenvalues}) is odd. The following table
lists the solution for several lowest eigenvalues (taking $b_0=1$ for even $l$ and $b_1=1$ for odd $l$).
\begin{eqnarray}
\qquad l&\qquad c_l\qquad&\qquad\phi_c(x)\nonumber\\[1mm]
0&2&1\nonumber\\
1&6&x\nonumber\\
2&12&1-5x^2\label{eq:eigenvaluescaseII}\\
3&20&x-\frac{7}{3}x^3 \nonumber\\
4&30&1-14x^2+21x^4 \nonumber\\
.&.&.\nonumber\\[-3mm]
.&.&.\nonumber\\[-3mm]
.&.&.\nonumber
\end{eqnarray}
In fact, up to a multiplicative constant, the functions $\phi_c(x)$ are
Gegenbauer polynomials \cite{redner_01,gra_ryz_94}.

The same procedure can be used for finding the eigenvectors of the
adjoint operator, solving Eq. (\ref{eq:forpsi}). 
We expand the function $\psi_c(x)$ in power series
\begin{equation}
\psi_c(x)=\sum_{l=0}^\infty d_l\,x^l
\label{eq:seriesforpsi}
\end{equation}
and find the recurrence relation for the coefficients
\begin{equation}
d_{l+2}      = \frac{(l-1)l-c}{(l+1)(l+2)}\,d_l\; .
\end{equation}
Again we conclude that the only acceptable values of $c$ are given by
condition $c=c_l\equiv (l+1)(l+2)$ for $l=0,1,2,...$ and in this case
the eigenvectors are polynomials of order $l+2$ in the variable
$x$. The following table lists the solution for lowest eigenvalues
(taking $d_0=1$ 
for even $l$ and $d_1=1$ for odd $l$). 
\begin{eqnarray}
\qquad l&\qquad c_l\qquad&\qquad\psi_c(x)\nonumber\\[1mm]
0&2&1-x^2\nonumber\\
1&6&x-x^3\nonumber\\
2&12&1-6x^2+5x^4\label{eq:lefteigenvaluescaseII}\\
3&20&x-\frac{10}{3}x^3+\frac{7}{3}x^5\nonumber\\
4&30&1-15x^2+35x^4-21x^6 \nonumber\\
.&.&.\nonumber\\[-3mm]
.&.&.\nonumber\\[-3mm]
.&.&.\nonumber
\end{eqnarray}
It is important to note that the set of right eigenvalues coincides
with the set of left eigenvalues, which proves consistency of our
approach.

Note that neither $\phi_c(x)$ nor $\psi_c(x)$ are orthogonal
polynomials. Instead, they are mutually orthogonal, i. e. 
\\
$\int_{-1}^{1}
\phi_c(x) \psi_{c'}(x)\mathrm{d}x=0$ for $c\ne c'$. This is due to
the fact that the operator $\mathcal{L}$ is not self-adjoint.
{}
\end{document}